\documentclass[a4paper,12pt]{article}
\usepackage{graphicx}%
\usepackage{amsmath}
\begin{document}
%
\title{ Search for $NN$-decoupled dibaryons using the process $pp \to \gamma \gamma X$
below the pion production threshold}
%
\author{ A.S.Khrykin, V.F.Boreiko, Yu.G.Budyashov, S.B.Gerasimov,\\
 N.V.Khomutov, Yu.G.Sobolev, V.P.Zorin}
%
\maketitle
\begin{abstract}
The energy spectrum for high
energy $\gamma$-rays ($E_\gamma \geq 10$ MeV) from the process $pp
\to \gamma \gamma X$ emitted at $90^0$ in the laboratory frame
has been measured at an energy below the pion
production threshold, namely, at 216 MeV. The resulting photon energy
spectrum extracted from $\gamma-\gamma$ coincidence events
consists of a narrow peak at a photon energy of
about 24 MeV and a relatively broad peak in the energy range of
(50 - 70) MeV. The statistical significances for the
narrow and broad peaks are 5.3$\sigma$ and 3.5$\sigma$, respectively.
This behavior of the photon energy
spectrum is interpreted as a signature of the exotic dibaryon
resonance $d^\star_1$ with a mass of about 1956 MeV which is assumed
to be formed in the radiative
process $pp \to \gamma \ d^\star_1$ followed by its electromagnetic decay
via the
$d^\star_1 \to pp \gamma$ mode.
The experimental spectrum is compared with those obtained by means
of Monte Carlo simulations.
\end{abstract}
\section{Introduction}
Direct-production of two hard photons in nucleon-nucleon
collisions ($NN\gamma\gamma$) at intermediate energies, unlike
production of single photons ($NN$ bremsstrahlung), is one of those
fundamental processes which are still poorly explored both
theoretically and experimentally. However, a study of this process
can yield new important information on the underlying mechanisms
of the $NN$ interaction that is complementary to that obtained from
investigations of other processes. In particular, the
$NN\gamma\gamma$ process can be used as a sensitive probe for
experimental verification of the possible existence of
$NN$-decoupled nonstrange dibaryon resonances. These are two-baryon
states $^2B$ with zero strangeness and exotic quantum numbers,
for which the strong decay
$^2B \to NN$ is either strictly forbidden by the Pauli
principle (for the states with the isospin $I=1(0)$ and with an even
sum $L+S+I$, where $L$ is the orbital momentum and $S$ is the spin),
or is strongly
suppressed by the isospin selection rules (for the states with
$I=2$)\cite{GerKh93,EGK95}. Such dibaryon
states cannot be simple bound systems
of two color-singlet nucleons, and
a proof of their existence would have consequences of fundamental
significance for the theory of strong interactions.
The $NN$-decoupled dibaryon states are predicted in a series
of $QCD$-inspired models\cite{QCDinm,Kondr,Kopel}.
Among the predicted dibaryons, there are those with masses both
below and above the pion production ($\pi NN$) threshold.
The $NN$-decoupled dibaryons with masses below the $\pi NN$ threshold
are of special interest,
since they may decay mainly into the $NN\gamma$ state, and
consequently, their widths should be very narrow ($\le 1 keV$).
Narrow dibaryon states have been searched for
in a number of experiments\cite{Tat00}, but none has provided
any convincing evidence for their existence.
Most of the past searches, however, were
limited to dibaryons coupled to the $NN$ channel, and,
to our knowledge, only a few were dedicated to the
$NN$-decoupled dibaryons\cite{Clement,Bilger,Konob}.
At the same time, since the dibaryons we are interested in do not couple to the
$NN$-channel, then, depending on their production and decay modes,
they may have escaped detection up to now and may naturally appear
in dedicated experiments.\\
If the $NN$-decoupled dibaryons
exist in nature, then the $NN\gamma\gamma$ process may proceed, at
least partly, through the mechanism that directly involves the
radiative excitation $NN \to \gamma \ ^2B$ and decay $^2B \to
\gamma NN$ modes of these states.
In $NN$ collisions at energies below the $\pi NN$
threshold, these production and decay modes of
the $NN$-decoupled dibaryon resonances with masses
$M_R \leq 2m_N + m_\pi$ would be unique or dominant.
The simplest and clear way of revealing them
is to measure the photon energy spectrum of the
reaction $NN\gamma\gamma$. The presence of an $NN$-decoupled dibaryon
resonance would reveal itself in this energy spectrum as a narrow line associated with
the formation of this resonance and a relatively
broad peak originating from its three-particle decay.
In the center-of-mass system, the position of the narrow line($E_R$)
is determined by the energy of colliding nucleons ($W=\sqrt{s}$)
and the mass of this dibaryon resonance as $E_R=(W^2-M_R^2)/2W$.
An essential feature of the two-photon production in $NN$
collisions at an energy below the pion production threshold is that,
apart from the resonant mechanism in question, there should only
be one more source of photon pairs.
This is the double $NN$-bremsstrahlung reaction.
But this reaction is expected to play a minor role.
Indeed, it involves two electromagnetic vertices, so that one may
expect that the $NN\gamma\gamma$-to-$NN\gamma$ cross section ratio
should be of the order of the fine structure constant $\alpha$.
However, the cross section for $NN\gamma$ is already
small (for example, the total $pp$-bremsstrahlung cross section at energies
of interest is a few $\mu b$).\\
The process $pp \to pp \gamma
\gamma$ at an energy below the pion production threshold provides
an unique possibility of searching for the $NN$-decoupled dibaryon
resonances in the mass region $M_R \leq 2m_p+m_\pi$ with
quantum numbers $I(J^P)=1(1^+,3^+,etc.$) or those with $I=2$
($J$ is the total spin, and $P$ is the parity of a dibaryon).
The preliminary experimental studies of the
reaction $pp \to \gamma\gamma X$ at an incident proton
energy of about 200 MeV\cite{Panic96,Menu97} showed that the photon energy
spectrum of this reaction had a peculiar structure ranging from
about 20 MeV to about 60 MeV. This structure was interpreted as
an indication of the possible existence of an $NN$-decoupled dibaryon
resonance(later called $d^\star_1$) that is produced
in the process $pp \to \gamma \ d^\star_1$ and subsequently decays via
the $d^\star_1 \to pp \gamma$ channel.
Unfortunately, a relatively
coarse energy resolution and low statistics did not allow
us to distinguish the narrow $\gamma$-peak associated with
the $d^\star_1$ production from the broad $\gamma$-peak due to the decay
of this resonance and, hence, to determine the resonance mass exactly.
To get a rough estimate of the $d^\star_1$ mass, we admitted that
the expected narrow $\gamma$-peak is positioned at the center of
the observed structure and thus obtained $M_R \sim 1920$ MeV.
However, if the uncertainty in the position of the narrow $\gamma$-peak
in the structure is taken into account,
the actual $d^\star_1$ mass might be considerably
different from such a crude estimate.
The possible existence of this dibaryon resonance was also probed in
the proton-proton bremsstrahlung data taken by
the WASA/Promice Collaboration at the CELSIUS accelerator\cite{WASA}
for proton energies of 200 and 310 MeV.
However, the trigger used in those experimental studies to select events
was designed and optimized for investigating the usual $pp \to pp\gamma$
bremsstrahlung. The selected events were those
corresponding to two simultaneously detected protons emerging in the forward
direction at angles between $4^0$ and $20^0$ with respect to the
beam axis, each of which had a kinetic energy exceeding the present
threshold (40 MeV). The analysis of those data resulted only in upper
limits on the dibaryon production cross section
in the mass range from 1900 to 1960 MeV.
Therefore, to clarify the situation with the dibaryon
resonance $d^\star_1$, we have decided to measure the energy spectrum
of the $pp \to pp\gamma\gamma$ reaction more carefully.
\section{Experimental details and event selection}
The experiment was performed using the variable energy proton
beam from the phasotron at the Joint Institute for Nuclear
Research(JINR).
The schematic layout of the experimental setup is
shown in Fig. 1.
\begin{figure}[t]
\begin{center}
\includegraphics[width=80 mm]{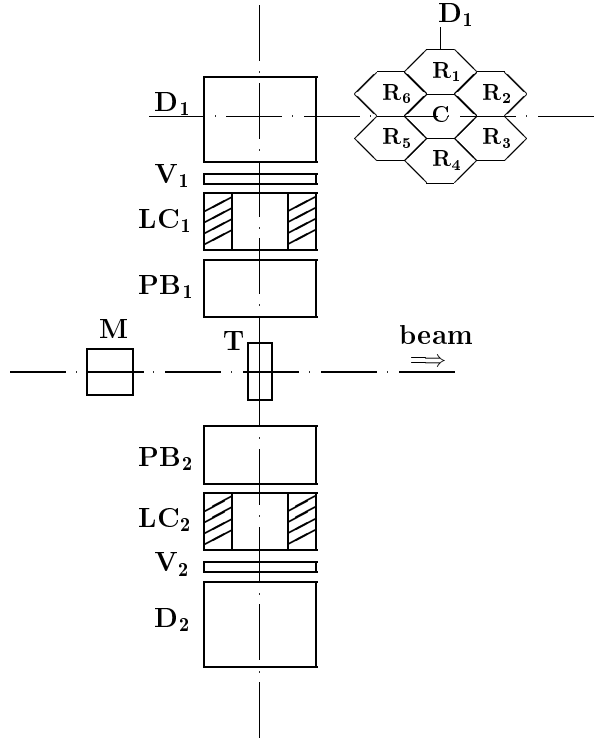}%
\end{center}
\caption{Experimental layout. Proton beam
approaches from the left.
$M$ is the monitor of the incident beam, $T$ is the target,
$D_1$ and $D_2$ are the $\gamma$-ray detectors, $V_1$ and $V_2$
are the plastic scintillators, $LC_1$ and $LC_2$ are the lead collimators,
 $PB_1$ and $PB_2$ are the polyethylene bars. At the top right of the
figure the cross section of the
$\gamma$-ray detector $D_1$ is shown.
$C$ is the central detector module and $R_1,R_2,..,R_6$ (ring) are the
detector modules surrounding the detector module $C$.}
\end{figure}
 The pulsed proton beam (bursts of about
7 ns FWHM separated by 70 ns) with an energy of about 216 MeV and
an energy spread of about 1.5\%
bombarded a liquid hydrogen target ($T$).
 The average beam intensity was on the order of $3.6 \cdot 10^8$ protons/s
and was monitored by an ionization chamber ($M$).
The liquid hydrogen target was a cylindrical cryogenic vessel with
a length of about 5 cm ($\sim 0.35 \ g/cm^2$) and a diameter of 4.5
cm, which had thin
entrance and exit windows made from 100 $\mu m$ thick mylar foil.
Both $\gamma$-quanta of the reaction $pp\gamma\gamma$ were detected by
two $\gamma$-ray detectors ($D_1$ and $D_2$) placed
in a horizontal plane, symmetrically on either side of the beam
at a laboratory angle of $90^0$ with respect
to the beam direction.
The $\gamma$-detector
$D_1$ was an array of seven individual detector modules (see Fig. 1):
the central detector module ($C$) and six detector modules ($R_1,
R_2,..,R_6$) surrounding the central detector module. Each
detector module was a 15 cm thick $CsI(Tl)$ crystal having a hexagonal
cross section with an outer diameter of 10 cm. The detector modules
$R_1,R_2,..,R_6$ (ring) were designed to measure the energy of
secondary particles and photons leaving the central detector module $C$.
Thus, the energy of each photon detected by the detector $D_1$ was
determined by summing up the energy deposited in the central
detector module and the energies deposited in the detector modules of
the ring.
Besides, the ring was used as an active shielding from cosmic ray
muons. The shielding was based on the fact that a cosmic ray
muon that passed through the detector $D_1$ should deposit an energy
at least greater than 50 MeV in the ring . The energy
resolution of a single $CsI(Tl)$ crystal was measured to be
about 13\% for $E_\gamma = 15.1$ MeV.
 The second $\gamma$-ray detector $D_2$ was a cylindrical
$NaI(Tl)$ crystal 15 cm in diameter and 10 cm in length.
Compared to the detector $D_1$, it had a poorer energy
resolution for $\gamma$-rays with energies $E_\gamma > 10$ MeV.
The detector $D_2$ was mainly designed to select $\gamma-\gamma$
coincidence events.
To reject events induced by charged particles, plastic scintillators ($V_1$
and $V_2$) were put in front of each $\gamma$-detector. Both
$\gamma$-ray detectors were placed in lead houses with walls 10 cm
thick which were in turn surrounded by a 5 cm borated
parafin shield. The front wall of every lead house had an opening
that served as a collimator ($LC_1$ and $LC_2$) limiting the angle acceptance of the
$\gamma$-detector.  The solid angles covered by
the detectors $D_1$ and $D_2$ were 43 msr and 76 msr, respectively.
To attenuate prompt particles coming from the target, 15 cm
polyethylene bars ($PB_1$ and $PB_2$) were placed in between the target and each
$\gamma$-detector.\\
The electronics associated with the
$\gamma$-detectors and the plastic scintillators provided amplitude
and time signals suitable for digitizing by an $ADC$ and a $TDC$.
The energy threshold for both the detector module $C$ and the
$\gamma$-detector $D_2$ was set at about 7 MeV.
The amplitude signals
from the $\gamma$-detectors were sent to the $ADCs$. The time signals from
the $\gamma$-detectors and the plastic scintillators were sent to the $TDCs$.
In addition, the radio-frequency (RF) signals from the accelerator
were also sent to the TDC.
Each of these signals was related with the time of arrival of the proton beam
burst at the target. Therefore, the time spectrum of the RF-signals
provided the means to select those events which correlate with proton bursts.
The time signal from
the detector module $C$ started
all the $TDCs$ and  opened the $ADC$ gates.
In parallel, the time signals from the detector module $C$ and the
$\gamma$-detector $D_2$ were sent to a coincidence circuit to select
those events
which had signals from both the $\gamma$-detectors within a 250 ns time interval.
The data acquisition system was triggered by signals from the
coincidence circuit and the corresponding events were recorded in the event-by-event mode
on the hard disk of the computer.
A further selection of $\gamma-\gamma$ coincidence
candidate events was done during off-line data processing.
The time interval of 250 ns was chosen
to record two different types of events, "prompt" and
"delayed". A prompt event is that in which both photons are due to
the same beam burst. Such events may be attributed to
either real $\gamma-\gamma$ events or to random coincidences.
A delayed event is that in which photons are due to two
different beam bursts, and, therefore, it must be
attributed to a random coincidence. This type of events
permits one to derive the component of random coincidences
mixed with real prompt $\gamma-\gamma$ events.

Data were taken for the target filled with liquid hydrogen
and for the empty one.
The off-line analysis of both data sets was done in exactly the same way.
The selection procedure for events associated with
the process $pp \to \gamma \gamma X$ included the following main
operations:
\begin{itemize}
\item[-]
Rejection of events induced by charge particles coming from the
target. For this purpose, all events having appropriate time
information from any plastic scintillators were rejected.
\item[-]
Rejection of events induced by cosmic ray muons.
These events were rejected by a requirement that in every event
the energy detected by the ring should not exceed 30 MeV. The
number of events induced by cosmic ray muons was found to be
insignificant.
\item[-]
Selection of events which are caused by proton beam bursts.
This procedure was done with the help of a time spectrum of
RF-signals. An example of this spectrum for a measurement with
the full target is shown in Fig. 2.
The time spectrum of RF-signals consisted of a peak of useful events
superimposed on a background of events due to random coincidences.
The time region confining the peak of useful events was
determined,
and events outside of this region were considered as a background
and were rejected.
\begin{figure}[th]
\begin{center}
\includegraphics[width=60 mm]{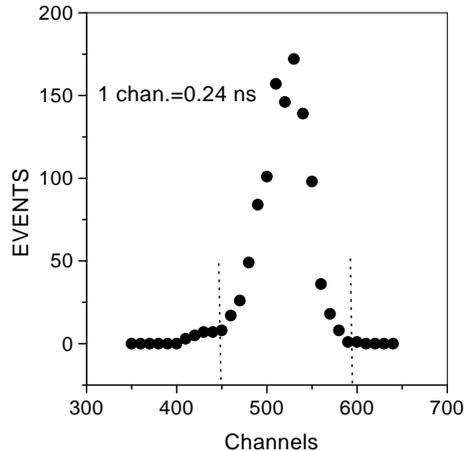}
\end{center}
\caption{\small{Time spectrum of RF-signals. The dashed lines
indicate the time interval confining the peak of useful events.}}
\end{figure}
\item[-]
Selection of events associated with $\gamma-\gamma$ coincidences.
To do this, a spectrum of time signals
from the detector $D_2$ (coincidence time spectrum) was used.
A part of this spectrum for a measurement with the full target
including the peak of prompt events and one of the peaks of delayed
events is shown in Fig. 3.
The coincidence time spectrum consisted of the peak of prompt events and a few
peaks of delayed events. The time region confining
the prompt peak was determined, and events associated with it
were considered as $\gamma-\gamma$ coincidence events and were
accepted.
\begin{figure}
\begin{center}
\includegraphics[width=60 mm]{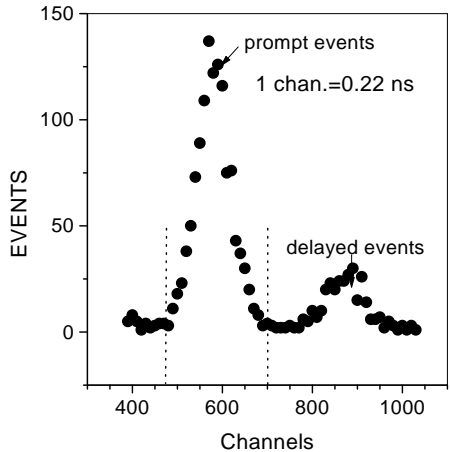}
\end{center}
\caption{\small{Coincidence time spectrum. The dashed lines
indicate the time interval confining the peak of prompt events.}}
\end{figure}
\end{itemize}
\section{Results and discussion}
 Data for the full and empty target were taken
in two successive runs for $\sim$31 h and $\sim$21 h, respectively.
The integrated luminosity of about $8.5\  pb^{-1}$ was
accumulated for the measurement with the full target.
The resulting energy spectra of prompt $\gamma-\gamma$ events as a function of the
photon energy measured by the detector $D_1$
for the full and the empty target are presented in Fig. 4.
Apart from real $\gamma-\gamma$ events, these spectra
also contain the contributions due to random
coincidences which were determined by using delayed events.
Such contributions to the full- and empty-target spectra
were found to be identical within the statistical errors.
\begin{figure}[t]
\includegraphics[width=80 mm]{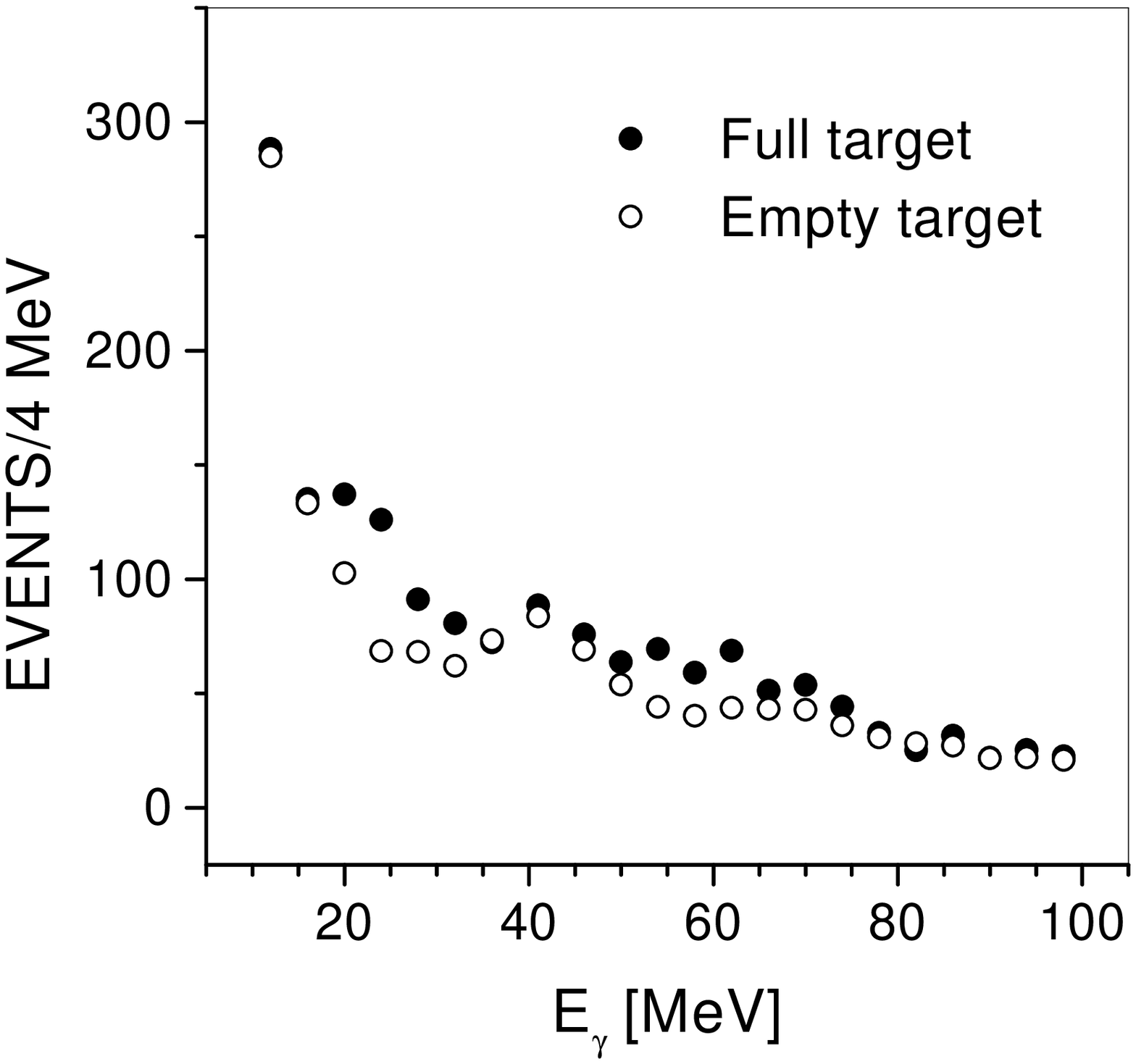}
\caption{\small{The photon energy spectra of prompt $\gamma-\gamma$ events
for data taken with the full target (filled circles) and
the empty target (open circles).}}
\end{figure}
The contribution of the random coincidence background to the
empty-target spectrum of prompt $\gamma-\gamma$ events was less than 20\%.
The behavior of the empty-target spectrum is determined by photon
pairs produced in collisions of protons with the mylar windows of the
target vessel.
Since carbon is the major component of mylar,
we also performed the measurement with the carbon target
to make sure that the observed shape of the empty-target spectrum
is due to the process $^{12}C(p,\gamma \gamma)X$.
It was found that the empty- and
carbon-target spectra have the same shape.

The photon energy spectrum of the process $pp \to \gamma \gamma X$
obtained after subtraction of the empty target
contribution is shown in Fig. 5.
As can be seen, this spectrum
consists of a narrow peak at a photon energy of about 24
MeV and a relatively broad peak in the energy range from about 50 to
about 70 MeV. The statistical significance
expressed in the number of standard
deviations (N.S.D.) was calculated for both the sharp and
the broad peak using the expression
$N.S.D.=(\sum_{i}1/{N^{eff}_i/{\sigma}_i}^2)/(\sum_{i}1/{{\sigma}_i}^2)^{1/2}$,
where $N^{eff}_i$ and ${\sigma}_i$ are the number of events and
the error bar at the energy $E_i$, respectively.
As a result, we found that the number of standard deviations
is equal to 5.3 for the narrow peak and to 3.5 for the broad peak.
The width (FWHM) of the narrow peak
was found to be about 8 MeV. This width is comparable with that
of the energy resolution of the experimental setup.
It is important to note that the number of events in the narrow peak
($\sim$ 130) is approximately equal to half the total number of
events in the spectrum ($\sim$ 250).
The observed behavior of the photon energy
spectrum agrees with a characteristic signature of the
sought dibaryon resonance $d^\star_1$ that is formed and decays
in the radiative process $pp \to \gamma \ d^\star_1 \to pp \gamma\gamma$.
In that case the narrow
peak should be attributed to the formation of
this dibaryon, while the broad peak should be assigned
to its three-particle decay. Using
the value for the energy of the narrow peak $E_R \sim 24$ MeV, we obtained
the $d^\star_1$ mass  $M_R \sim 1956$ MeV. The uncertainty of the
given value is primarily due to the uncertainties in the value of the
proton beam momentum and the energy of the narrow $\gamma$-peak.
We estimate the total uncertainty in the $d^\star_1$ mass to be $\pm 6$
MeV. In such a way we obtain $M_R = 1956 \pm 6$ MeV. This result is
inconsistent with the rough estimate of the $d^\star_1$
mass $M_R \sim 1920$ MeV we obtained earlier\cite{Panic96,Menu97}.
The results of the present experiment enable us to estimate the differential
cross section of the resonance production of two photons in the process
$pp \to \gamma \ d^\star_1 \to pp \gamma\gamma$ by the formula:
\begin{equation}
\frac{d^2 \sigma}{d \Omega_1 d \Omega_2}
 = \frac{N^{eff}}{L \Delta \Omega_1 \Delta \Omega_2},
\end{equation}
where $N^{eff}$ is the total number of events observed, $L$ is
the luminosity, and $\Delta \Omega_1$ and $\Delta \Omega_2$ are the solid angles
covered by the $\gamma$-ray detectors. We found that at an energy
of 216 MeV this cross section for photons emitted symmetrically
at $\theta_{lab} = \pm 90^0$ is $\sim$ 9 nb/sr$^2$.

Having assumed that the $pp \to \gamma \ d^\star_1 \to pp
\gamma\gamma$ process with the $d^\star_1$ mass of 1956 MeV is the only
mechanism of the reaction $pp \to pp \gamma\gamma$, we calculated
the photon energy spectra of this reaction for a proton energy of
216 MeV. It was also assumed that the
radiative decay of the $d^\star_1$ is a dipole $E1(M1)$ transition
from the two-baryon resonance state to a $pp$-state in the continuum.
The calculations were carried out with the help of Monte Carlo
simulations which included the geometry and the energy resolution of
the actual experimental setup. The photon energy spectra were
calculated for two different scenarios of the $d^\star_1$ decay.
The difference between them was that one of these scenarios took into
account the final state interaction ($FSI$) of two outgoing
protons whereas in the other that interaction was switched off.
Each of the
scenarios imposed some restrictions on possible quantum numbers of
the dibaryon state in question. The scenario including the $FSI$
implies that the final $pp$-system is in the singlet $^1S_0$
state and consequently it should take place, in particular,
for the isovector
$1^+$ dibaryon state (the simplest exotic quantum numbers), namely,
$1^+ \stackrel{M1}{\longrightarrow} 0^+$. Moreover such a scenario
can take place for any isotensor dibaryon state with the exception
of the $0^+$ or $0^-$ state.
 At the same time, the scenario
in which the $FSI$ is switched off, is most likely to occur for
the isotensor $0^\pm$ dibaryon state.
The spectra calculated for these two decay scenarios
 and normalized to the total number of $\gamma-\gamma$
events observed in the present experiment are shown in Fig.5.
\begin{figure}[t]
\begin{center}
\includegraphics[width=80 mm]{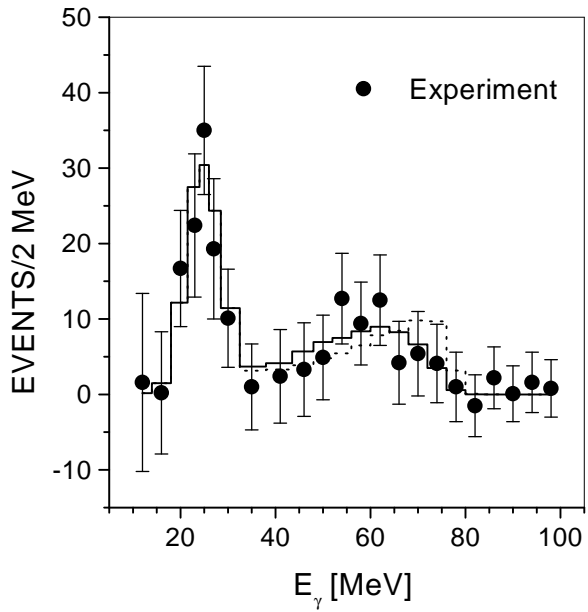}
\end{center}
\caption{\small{Experimentally observed
energy spectrum for photons from the $pp\gamma\gamma$
process and energy spectra for photons from the process
$pp \to \gamma d^\star_1 \to \gamma\gamma pp$
calculated with the help of Monte Carlo simulations for two $d^\star_1$
decay scenarios: without the FSI(solid line)
and with the FSI(dashed line). The solid and dashed lines are overlapped
in the energy ranges $E_\gamma \leq 30$ MeV and $E_\gamma \geq 80$ MeV.}}
\end{figure}
Comparison of these spectra with the experimental spectrum indicates
that both the calculated spectra are in reasonable agreement with
the experimental one within experimental uncertainties.
In other words, the statistics of the
experiment is insufficient to draw any firm conclusions in favor
of one of these scenarios and thereby to limit possible
quantum numbers of the observed dibaryon state. In this connection, we
would like to emphasize that the experimental data on the
process $\pi^- d \to \gamma\gamma X$ can shed some light on the
properties of the dibaryon state we have observed. The point is
that the $d^\star_1$ dibaryon can in principle be excited in
this process via the mode $\pi^- d \to \gamma d^\star_1 \to \gamma\gamma nn$.
However, excitation of the dibaryon $d^{\ast}_{1}$ with isospin
$I=1$ is allowed by the isospin selection rules, whereas excitation of
the isotensor $(I=2)$ resonance is expected to be strongly
suppressed\cite{Ge97,Ge98}. Therefore, the preliminary results of the study of the
reaction $\pi^- d \to \gamma\gamma X$ carried out by the RMC
Collaboration at TRIUMF\cite{Triumf} indicating no
enhancement of the two-photon yield in the process $(\pi^{-} d)_{atom}
\to \gamma \gamma nn$ compared to the usual one-nucleon
mechanism $\pi^{-}p \to \gamma\gamma n$ on the proton bound in the deuteron
may imply that the $d^{\ast}_{1}$ dibaryon has isospin 2.

Finally, as we have already mentioned in the Introduction, the presence of the
$pp \to \gamma \ d^\star_1(1920) \to \gamma\gamma pp$ process was
sought in the Uppsala $pp$-bremsstrahlung data\cite{WASA}.
Based on the detailed data analysis the authors\cite{WASA}
conclude that they have found no evidence for a dibaryon in the mass range
from 1900 to 1960 MeV.
There is, however, one important feature of the process in
question which was omitted in Ref.\cite{WASA}.
It is related to the angular distribution of
protons from the dibaryon decay $d^\star_1 \to pp\gamma$.
The authors\cite{WASA} assumed that such a distribution
is isotropic in the center-of-mass system of the reaction. However,
that is not the case.
To understand the angular distribution of the protons
from the process $pp \to \gamma \ d^\star_1(1956) \to \gamma\gamma pp$
we performed phase space calculations using the
CERN code\cite{FOWL}.
The calculations were done for two possible decay modes of the
dibaryon. One is the dibaryon
decay to the singlet $^1S_0$ two-proton state ($^1S_0\{pp\}$).
It is allowed, for instance, for $d^\star_1$ with
quantum numbers $1(1^+)$ or $2(1^\pm)$.
The second implies that the dibaryon decays to any other
two-proton state.
It was also assumed that the $\gamma$-radiation
from the given process has a dipole character.
The results of the calculations (in the laboratory) for an incident energy of 310 MeV and
for the case when the polar angle of observed photons
was constrained to the range between $30^0$ and $90^0$
($\gamma$-ray detector arrangement in the Uppsala experiment)
are shown in Fig.6.
The same calculations done at 200 MeV led to similar results.
\begin{figure}[t]
\begin{center}
\includegraphics[width=80 mm]{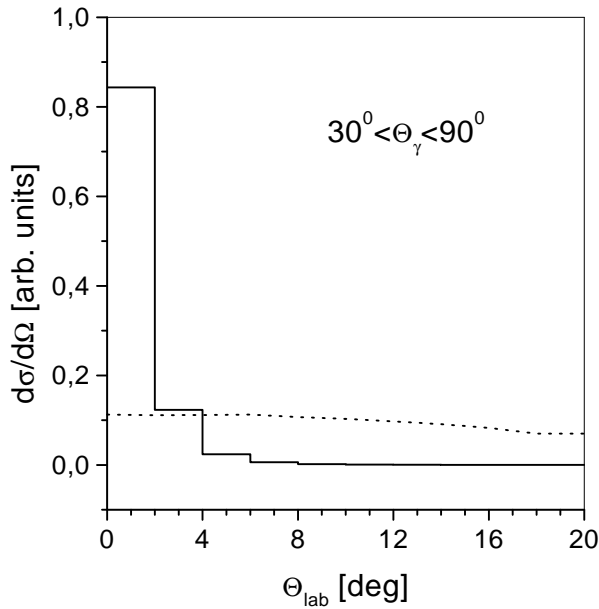}
\end{center}
\caption{\small{The angular distribution of the protons
(in the laboratory) from the process
$pp \to \gamma \ d^\star_1(1956) \to \gamma\gamma pp$
calculated with FOWL\cite{FOWL} for the case when the polar angle
of observed photons is constrained to the range between $30^0$ and $90^0$.
The solid line corresponds to four-body phase space with $^1S_0$ FSI and the
dashed line to that without the FSI between the outgoing protons.}}
\end{figure}
As can be seen from Fig.6, the protons from the process
$pp \to \gamma \ d^\star_1(1956) \to \gamma\gamma \ {^1S_0\{pp\}}$ are mainly concentrated in a narrow angular cone
at $\theta_{lab} \le 4^0$.
Since in the Uppsala experiment events were only accepted if
the polar angle of each proton fell within a forward
cone between $4^0$ and $20^0$, the dominant portion of
events associated with the given process
was beyond this angular range and, hence, were not
present in the analyzed data\cite{WASA}. In other words, if we assume
that the $d^\star_1(1956)$ has quantum numbers such as $1(1^+)$ or
$2(1^\pm)$, then these data can contain only such events from the process
$pp \to \gamma \ d^\star_1(1956) \to \gamma\gamma \ g{^1S_0\{pp\}}$
in which both final protons are associated with the tail of the angular
distribution presented in Fig.6. Our naive estimations have shown that the
fraction of these events in the data\cite{WASA} for 200 MeV and 310
MeV with respect to the corresponding total number of events from
the process of interest are approximately smaller than 20\% and
2\%, respectively. These estimates together with the differential
cross section ($\frac{d^2 \sigma}{d \Omega_1 d \Omega_2})$
obtained in the present experiment permitted us, in
turn, to estimate the expected number of events in the Uppsala
experiment. The corresponding values for the measurements at 200
MeV and 310 MeV are equal to $\sim$2 and $\sim$130, respectively. It was
assumed in those calculations that the energy dependence of the
cross section is $E_R^3$ (dipole radiation), where $E_R=(s-M_R^2)/2
\sqrt s$. Our estimate for 200 MeV is in agreement with the
corresponding null result obtained in the Uppsala measurements. As
for the result at 310 MeV, we would like to note the following.
First, the experimental sensitivity of the data\cite{WASA} at 310 MeV
to the process in question was limited by a large
 background stemming from $\pi^0$ decays.
We found that cuts applied in Ref.\cite{WASA} to remove this background
reduced the number of dibaryon candidates observed by a factor
of $\sim$3. Second, the presence of a narrow
dibaryon should appear in the $pp\gamma$ invariant mass spectrum
as a sharp peak and a wider one (mirror) that arises because of the
$pp\gamma$ invariant mass calculation with a photon originating
from a dibaryon production.
The authors\cite{WASA} believed that inside a range of dibaryon
masses from 1900 to 1960 MeV these peaks would be well
separated from one another. Therefore, as a signal for dibaryon
production they hoped to observe a sharp peak $\sim$5 MeV wide.
To find the invariant mass distribution of $pp\gamma$ events from
the $pp \to \gamma \ d^\star_1(1956) \to \gamma\gamma pp$ process
at 310 MeV we performed a Monte Carlo simulation of the WASA/PROMICE experimental setup.
It appeared that a broad mirror and a sharp peak
merge in this distribution and it consists of one wide peak
extending from $\sim$1943 to $\sim$1968 MeV. We think that in Ref.\cite{WASA}
such a peak could be overlooked. Of course, the estimates obtained for
the expected number of events in the Uppsala experiment
are very approximate. To make more precise estimates it is desirable
to have a consistent description of the
$pp \to \gamma d^\star_1 \to \gamma\gamma pp$ process.
\section{Conclusion}
 The $\gamma$-ray energy spectrum for the $pp \to \gamma \gamma X$ reaction
at a proton energy below the pion production threshold has been measured
for the first time.
The spectrum measured at an energy of about 216 MeV
for coincident photons emitted at an angle of $90^0$ in the laboratory frame
clearly evidences the existence of the $NN$-decoupled dibaryon
resonance $d^\star_1$ with a mass of $1956 \pm6$ MeV that is
formed and decays in the process $pp \to \gamma \ d^\star_1 \to pp \gamma\gamma$.
The data we have obtained, however, are still incomplete, and
additional careful studies of the reaction $pp \to pp \gamma \gamma$
are needed to get proper parameters (mass, width, spin, etc.)
of the observed dibaryon state.

\end{document}